%% file: holte-nadathur.tex
\documentclass{llncs}
\usepackage{eepic}
\usepackage{epic}
\usepackage{latexsym}

\input macros

\usepackage{makeidx}  % allows for indexgeneration

\begin{document}

\pagestyle{plain}

\mainmatter 

\title{Modularity and Separate Compilation\\ in Logic Programming
\vspace{-0.4cm}}

\titlerunning{Modularity and Separate Compilation}

\author{Steven Holte and Gopalan Nadathur}

\institute{Department of Computer Science and Engineering,
University of Minnesota,\\
4-192 EE/CS Building, 200 Union Street SE, Minneapolis, MN 55455\\
\email{Email: \{holte,gopalan\}@cs.umn.edu, Fax: 612-625-0572}\\
\ignore { Home Page: \texttt{http://www.cs.umn.edu/\homedir gopalan}}
}

\maketitle 

\vspace{-0.5cm}

\begin{abstract}
The ability to compose code in a modular fashion is important to
the construction of large programs. 
In the logic programming setting, it is desirable that such
capabilities be realized through logic-based devices. 
We describe here an approach for doing this that is a simplification
and a rationalization of features previously supported in the {\it
  Teyjus} 
implementation of the $\lambda$Prolog language. 
Within this scheme, a module corresponds to a block of code whose
external view is mediated by a signature.  
Thus, signatures impose a form of hiding that is explained, at the
logical level, via existential quantifications over predicate,
function and constant names.  
Module interaction is based on {\it accumulation}, a mechanism
that translates into conjoining the clause definitions in them while 
respecting the scopes of existential quantifiers arising from  
signature interpretation.
Our first contribution is to show that this simple device for
statically structuring name spaces suffices for realizing features 
related to code scoping for which the dynamic control of predicate
definitions was earlier considered necessary. 
The module capabilities we present have previously been implemented in
the {\it Teyjus} system via the compile-time inlining of accumulated
modules. 
This approach has the drawback of not supporting separate
compilation. 
Our second contribution is a scheme that allows each distinct module
to be compiled separately, moving the task of inlining to a later,
complementary linking process.
\end{abstract}

\vspace{-0.5cm}

\input intro

%\vspace{-0.15cm}
\input core-lang
%\vspace{-0.15cm}
\input modularity-basis
%\vspace{-0.15cm}
\input modules-practical
%\vspace{-0.15cm}
\input sep-compilation
%\vspace{-0.15cm}
\input conc

%\vspace{-0.15cm}
\input ack

\input{holte-nadathur.bbl}
\end{document}

%% file: macros.tex
\long\def\ignore#1{}

\newcommand{\eg}{{\em e.g.}}

%% file: intro.tex
\vspace{-0.5cm}

\section{Introduction}\label{sec:intro}

\vspace{-0.25cm}

We are concerned in this paper with a treatment of modularity in logic
programming. Support for this feature is important to significant 
applications of the paradigm. The ability to develop a system
through the composition of small, well-defined units of code is
central to managing complexity and also facilitates
the reuse of programming effort. Moreover, modular development
installs boundaries in programs that can be important to the practical 
use of static analysis techniques and that are fundamental to the
notion of separate compilation and testing. In light of these facts,
it is not surprising that modularity aspects as they pertain to logic
programming have received theoretical treatment
\cite{HP98jlc,Mil89jlp,MP89,SW92}, have been included 
in practical systems \cite{ciao97,sicstus,quintus} and have been the
topic of  standardization deliberations concerning Prolog
\cite{ISO-Prolog-Modules}.  

Modularity notions have typically been incorporated into logic
programming systems by going outside the logical base
and introducing metalinguistic mechanisms for composing separately
constructed program fragments. This approach is a little unfortunate:
a strength of logic programming is its basis in 
logic that, with proper choices, can also be used to reason about
interactions between units of code \cite{Miller93elp}.
\ignore{
\footnote{Logic
can also be in informing particular choices such as those between
``atom-based'' and ``predicate-based'' approaches to hiding as we see
in this paper.}}
However, there is also a danger in focussing too heavily on just the 
logical aspects. Early logic-based approaches to controlling code
availability have, for instance, utilized 
the idea of implications or contexts in goals \cite{Mil89jlp,MP89}. These
mechanisms call for a dynamic management of predicate definitions and,
as such, their implementation is costly. Moreover, the
particular way in which context is handled leads sometimes to 
program behaviour that is counter to the practical understanding
of modularity. 

We describe a treatment of modularity in this paper that balances
logical and pragmatic considerations. This treatment draws on 
experience with a logic based approach \cite{Miller93elp}
employed in the {\it Teyjus} implementation \cite{NM99cade} of the
language $\lambda$Prolog. However, our ideas are quite general and can
be used in any logic programming setting that correctly implements
sequences of alternating existential and universal quantifiers in
goals.\footnote{Such a capability can be added to common logic
programming languages by changing the unification computation 
as indicated later.}
The devices we use are, in fact, surprisingly 
simple at a logical level. To support information hiding, we utilize
existential quantification over names. 
\ignore{; since such names could be those of
predicates and functions in addition to (first-order) constants, the
appropriate context for our ideas is that of a higher-order logic.}
Pragmatically, the hiding of names is effected by associating a 
signature with each module of code; all the names used in the module
and not appearing in the signature are treated as being implicitly
existentially quantified.  The composition of units of code,
accomplished via a mechanism known as {\it accumulation}, translates into the
conjoining of formulas. This leads to a {\it statically determined} 
code space but one in which the availability of predicate definitions
can be controlled by appropriately scoped existential
quantifiers. We show that these simple devices suffice for realizing
features such as scoping of predicate definitions,
data abstraction and module parameterization for which more
complicated dynamic code structuring capabilities were previously
thought to be necessary \cite{Mil89jlp,MP89}. A noteworthy point is
that the often problematic aspect of higher-order programming coexists
{\it completely naturally} with this approach to modularity.  
From a implementation perspective, accumulation can be treated through
a compile-time inlining of code \cite{NT99jflp}. 
However, true modularity requires support for separate compilation.   
Our second contribution consists of showing that this can be provided.  
In particular, we describe a scheme that 
permits each module to be compiled separately with the inlining
function being relegated to a later, link-time process. Although we do
not explicitly demonstrate it here, this two phase process
cumulatively expends effort similar to the compile-time inlining
method and it produces an identical executable image.

The rest of the paper is structured as follows. In the next section we
introduce a logical language that includes all the features needed to
capture our treatment of modularity. In
Section~\ref{sec:modularity-basis} we describe the main components of
the modules language and indicate their translation into the logical
core. The following section discusses usage paradigms and also
some practical embellishments to the basic modules language.
\ignore{
The following
section describes embellishments in the form of annotations or
compiler directives to the basic modules language; these directives do
not affect the translation into logic but can be used to control
acceptability of programs in pragmatically meaningful ways and may
also lead to more efficient implementations. }
In Section~\ref{sec:sep-compilation} we consider the issue of separate
compilation. Section~\ref{sec:conc} concludes the paper by contrasting
its contents with related work.

%% file: core-lang.tex
\vspace{-0.35cm}

\section{The Underlying Logical Language}\label{sec:core-lang}

\vspace{-0.3cm}

Our approach to information hiding involves the use of existential
quantification. Since we desire the ability to hide the names of
predicates and functions in addition to (first-order) constants, the
right context for our ideas is that of a higher-order
logic. We shall assume also that our language is typed although our
main ideas apply equally in an untyped setting. 

%The particular logical language that we use here is the one
%underlying $\lambda$Prolog. This choice is not essential: our main
%ideas apply equally to an (untyped) higher-order language that
%suitably generalizes Prolog.   

%Our logic, then, is both typed and higher-order. 
%that is a fragment of the 
%intuitionistic version of Church's Simple Theory of Types
%\cite{Church40}, 
%is typed and higher-order. 
We work with a set of types that initially contains 
{\it int}, {\it real}, {\it string} and {\it o}, those that can be
formed using the unary type constructor {\it list} and, finally, all
the function types, written as $\alpha \rightarrow \beta$, that can be
formed  from these types. The type {\it o} is that of
propositions. Type variables, denoted by tokens starting with
uppercase letters, are introduced as a shorthand for an infinite
collection of instance types.
Terms in the language are constructed from collections of typed 
constant and variable symbols using the operations of application and
abstraction. 
The initial set of constants includes the logical symbols {\it true}
denoting the always true proposition, $\supset$, $\lor$ and $\land$
denoting infix forms of implication, disjunction and conjunction and
two `schema' constants {\it sigma} and {\it pi} of type $(A
\rightarrow o) \rightarrow o$. The last two (family of) constants
represent generalized existential and universal quantifiers: $\exists
x\, P(x)$ and $\forall x\, P(x)$ are rendered in this logic as
$(sigma\ \lambda x P(x))$ and $(pi\ \lambda x P(x))$
respectively.\footnote{We write $P(x)$ to denote a term that possibly
has free occurrences of $x$.} We also assume the usual constants
denoting integers, reals and strings  
% (whose specific structures we do not describe here) 
and the schema
constants {\it   nil} of type {\it list~A} and the infix \verb+::+ of
type $A \rightarrow (list\ A) \rightarrow (list\ A)$ that provide for
a `builtin' notion of lists. 
Application is assumed to be left
associative. This leads to a curried notation for terms. If $p$
is an $n$-ary relation symbol, the expression 
$(p\ t_1\ \ldots\ t_n)$ denotes this relation between the terms
$t_1,\ldots,t_n$. Such an expression constitutes an atomic 
formula if its `head' $p$ is either {\it true} or is not a logical
constant. 
\ignore{
The initial set of constants also includes those
given by  sequences of digits denoting integers, by two sequences of
digits separated by a period denoting reals, by sequences of characters
enclosed between double quotes denoting strings and the schema
constants {\it   nil} of type {\it list~A} and the infix \verb+::+ of
type $A \rightarrow (list\ A) \rightarrow (list\ A)$ that provide for
a `builtin' notion of lists. 
}

In any real application it is necessary to introduce new sorts and
type constructors as well as new nonlogical constants. We describe
mechanisms in the next section for declaring such symbols. 
\ignore{As we shall
see there, modules provide a natural means for delineating the scope
of such declarations.} Types must also be associated with
variables. Such type information can be filled in by an inference
process and the language syntax may also allow it to be explicitly
provided.  

The logic that we shall use to expose our ideas is slightly expanded
version of the higher-order theory of Horn clauses. The central
components of this logic are the $G$ and $E$ formulas 
identified by the following syntax rules:
%\vspace{-0.5cm}
\begin{tabbing}
\qquad\=D\quad\=::=\quad\=\kill
\>$G$\>::=\>$true \ \vert\  A \ \vert\  G \lor G \ \vert\
G \land G \ \vert\  \exists x\,G $\\
\>$D$\>::=\>$A_r \ \vert\  G \supset A_r \ \vert\  \forall x\,D\
\vert\  D \land D$\\
\>$E$\>::=\>$D \ \vert\ E \land E\ \vert\ \exists x\,E$
\end{tabbing}
%\vspace{-0.5cm}
\noindent $A$ denotes atomic formulas here and $A_r$ 
represents atomic formulas whose heads are either constants distinct
from {\it true} or are existentially quantified in the enclosing
context. 
\ignore{Existential quantifiers in $E$ formulas are treated as
constants in the computation model described below. From this it
becomes clear that the head of an $A_r$ formula is always a nonlogical 
constant, albeit of smaller or larger scope.} A $D$ formula of the form
$A_r$ or $G \supset A_r$ is said to have $A_r$ as its head and its
body is either empty or $G$ depending on the case in
question. Such a formula is, as usual, part
of the definition of a predicate whose name is the head of the $A_r$
formula; this interpretation is meaningful because the computation model
described next ensures that this head is eventually always a
nonlogical constant. Restricted to the first-order setting, our
$D$ formulas are essentially typed versions of definite Horn
clauses or program clauses that underlie Prolog. 

Computation in the core language is oriented around trying to solve a
goal given by a $G$ formula given a collection of closed $E$
formulas that defines a program. Semantically, this notion is explained via
provability in intuitionistic logic. At a procedural level, this
translates into carrying out the following steps.\footnote{This
description should be read as a {\it precise} but {\it abbreviated}
rendition necessitated by paucity of space of an alternative
presentation via a transition system.} First, a signature is
determined by the (global) constants appearing in the program and the
goal and is used to determine instantiations for the free variables in
the goal. Then, the existential quantifiers at the front of $E$
formulas are instantiated by new constants, the available signature is
expanded to include these constants and conjunctions are eliminated to
transform the program to a set of $D$ formulas of the form $A_r$ or $G
\supset A_r$. Next, a complex goal is reduced recursively based on its
structure: a conjunctive goal results in an attempt to solve each
conjunct from the same program, a disjunctive goal becomes an attempt
to solve one of the disjuncts and an existential goal is instantiated
by a chosen closed term based on the current signature and not
containing $pi$ or $\supset$. Eventually, the goal encountered must be
atomic. This goal is solved trivially if it is {\it true}. If not, it
is matched with the head of a closed instance of a $D$ formula in the
program, resulting in a solution if the body is empty and the body of
the $D$ formula as a new goal to be processed in an identical fashion
otherwise.

The framework for computation described above possesses an interesting
ability to treat scope. Specifically, existentially quantified
variables in programs correspond to names or constants and explicit
quantification makes it possible to indicate a scope for such
names. Thus, in a program that contains the  formulas $\exists
x\,D_1(x)$ and $\exists x\,D_2(x)$, the `names' $x$ 
that appear in the two contexts are to be thought of as {\it distinct}
constants. This situation is to be contrasted with one where the
program contains the formulas $D_1(x)$ and $D_2(x)$, assuming that the
two $x$s meet the syntactic criteria for being treated as 
constants. Existential quantification also has an impact on the
availability of names in an outside context. Thus, the constant $x$ is
allowed to appear in instantiations of the free variable $y$ when an
attempt is made to solve the goal $G(y)$ from a program containing the
formula $D_1(x)$. However, this constant {\it may not} be so used if
the program formula is changed to $\exists x\,D_1(x)$ instead.

\ignore{
A substantive fact about the logic being considered is that the
procedure we have outlined for solving a $G$ formula from a set of $E$
formulas exactly captures provability in intuitionistic logic.
}

%% file: modularity-basis.tex
\vspace{-0.25cm}

\section{The Modules Language and Its Logical
  Interpretation}\label{sec:modularity-basis} 

\vspace{-0.25cm}

In developing real programs, it is necessary, first of all, to
identify a vocabulary of types and term constants. New type
constructors are defined in our language through 
{\it kind} declarations that have the form 
\vspace{-0.15cm}
\begin{tabbing}
\qquad\=\kill
\>\verb+kind  tyc1, ..., tycn  type -> ... -> type.+
\end{tabbing}
\vspace{-0.15cm}
where the arity of the constructors \verb+tyc1+,$\ldots$,\verb+tycn+
is one less than the number of occurrences of \verb+type+ in the
declaration. Term constants are identified through {\it type} declarations of 
the form 
\vspace{-0.15cm}
\begin{tabbing}
\qquad\=\kill
\>\verb+type  c1,...,cn  <type expression>.+
\end{tabbing}
\vspace{-0.15cm}
where the type expression is constructed using the available type
constructors.  

The notion of scope for kind and type declarations is important in
constructing type and predicate definitions. Modules impart a
structure to this space of names. Formally, a module begins with a
declaration of the form 
\vspace{-0.15cm}
\begin{tabbing}
\qquad\=\kill
\>\verb+module <name>.+
\end{tabbing}
\vspace{-0.25cm}
and continues with the kind, type and predicate definitions to be
associated with the indicated name. We adopt a file oriented view
of modules here: all the code defining a module named \verb+foo+
is to be found in a file with the name \verb+foo.mod+. Now, the 
boundaries of a module determine a textual notion of scope in that the
kind and type declarations that appear within them are interpreted as
ranging over all the other declarations contained in the
module. Consistent with this viewpoint, these boundaries also provide a
delimiting region for analyses that a compiler might perform in the
course of translating a source language program.  

The following definition of a module called \verb+store+ illustrates
module syntax:
\vspace{-0.25cm}
\begin{tabbing}
\qquad\=\kill
\>\verb+module  store.+\\
\>\verb+kind  store  type -> type.+\\
\>\verb+type  emp  (store A).+\\
\>\verb+type  stk  A -> (store A) -> (store A).+\\
\>\verb+type  init  (store A) -> o.+\\
\>\verb+type  add, remove  A -> (store A) -> (store A) -> o.+\\
\>\verb+init emp.+\\
\>\verb+add X S (stk X S).+\\
\>\verb+remove X (stk X S) S.+
\end{tabbing}
\vspace{-0.15cm}
This module identifies a representation for stores with three
associated operations, one for initializing a store and two others for
adding an element to and removing an element from an existing
store. Towards providing a definition that is parametric in the type
of the elements stored, this module declares a unary type constructor
for stores that is then used in the types of constants that implement
store representations. The particular realization of a store embedded
in this code is based on the idea of a stack. The last three lines in
the module are $D$ formulas that define the desired operations based
on this interpretation. The tokens that begin with uppercase letters 
stand, as usual, for variables that are implicitly universally
quantified at the head of the formula. 

Not all the declarations in a module are typically intended to be
externally visible. With the module \verb+store+, for example, it is
sensible to hide the actual representation of stores, requiring these
to be manipulated opaquely through the predicates \verb+init+,
\verb+add+ and \verb+remove+. In our language, the contents of a
module that are to be visible to the outside must be explicitly
identified through a signature that shares its name with the
module. Formally, a signature begins with a declaration of the form  
\vspace{-0.25cm}
\begin{tabbing}
\qquad\=\kill
\>\verb+sig <name>.+
\end{tabbing}
\vspace{-0.25cm}
and continues with the kind and type declarations to be associated
with the specified name. Thus, the following declarations constitute a  
signature that imposes the kind of view desired on the module \verb+store+:
\vspace{-0.25cm}
\begin{tabbing}
\qquad\=\kill
\>\verb+sig  store.+\\
\>\verb+kind  store  type -> type.+\\
\>\verb+type  init  (store A) -> o.+\\
\>\verb+type  add  A -> (store A) -> (store A) -> o.+\\
\>\verb+type  remove  A -> (store A) -> (store A) -> o.+
\end{tabbing}
\vspace{-0.1cm}

There is an obvious consistency requirement with signatures: they 
must identify all the type constructors that are needed
to sensibly interpret the types that appear in them. A syntactically
well-formed signature actually functions as an interface definition
for a module: a compiler must treat this as
a complete description of all the type and kind declarations emanating
from the module in any context where the module is used and it
must correspondingly check that these declarations agree with what is 
actually present in the module. In support of this kind of type
checking role, we shall adopt a file oriented view of signatures
similar to that for modules: the definition of the signature named 
\verb+foo+ is to be found in a file called \verb+foo.sig+.

We have thus far considered the {\it static semantics} of modules and
signatures. The {\it dynamic semantics} is explained by a translation
into an $E$ formula and a subsequent use of the computation model for
the core language.  This translation is actually quite simple: the
entire module is to be thought of as a, perhaps large, $E$ formula
that is obtained by conjoining all the $D$ formulas contained in it
and then existentially quantifying over all the local constants---{\it
i.e.} the constants not mentioned in the signature---in this
formula. Under this approach, the logical essence of the module
\verb+store+, for instance, is reduced to the formula 
\vspace{-0.15cm}
\begin{tabbing}
\qquad\=$\exists Emp \exists Stk ($\=\kill
\>$\exists Emp \exists Stk ($\>$(init\ Emp)\ \land$\\
\>\> $\forall X \forall S\, (add \ X \ S\ (Stk\ X \ S))\ \land$\\
\>\> $\forall X \forall S\, (remove\ X \ (Stk\ X\ S)\ S)).$
\end{tabbing}

The simplest form of module use occurs at the interaction level. In
particular, the attempt to solve a goal at the top level is always
made relative to a chosen module. In keeping with the preceding
discussion, from a lexical perspective, the
significance of such a relativization is that all the kind and type
declarations in the signature of the module become available in 
analyzing the syntax of the goal. The
computational significance of the relativization, on the other hand,
is that the goal is to be solved using only the definitions of global
predicates in the module and assuming an access at most to its global
constants. This interpretation actually implies a strong form of
hiding. Specifically, suppose that we are trying to solve the goal $G$ from the
module $M$. It is obvious that $G$ cannot refer directly to any
constant local to $M$. A little less obvious is the fact that
these local constants cannot also percolate out of $M$ in the form of
computation results. Thus, suppose that $G$ is a goal whose
only free variable is $y$. The results computed for this query are
then the instantiations for $y$ that lead to successful solutions of
$G$. The translation semantics for modules and the associated
operational semantics now clearly preclude
the appearance of the local  constants of $M$ in instantiations for
$y$.\footnote{An issue 
different from semantics is the efficient treatment of such
constraints on instantiations. We have shown elsewhere how to do this  
by assigning numeric labels to logic variables and constants and
by using these labels in unification \cite{Nad92int}.}

An important aspect of a module system is the mechanisms it provides
for realizing interactions between units of code. In our system, this
ability is obtained from a single operation referred to as {\it
  module accumulation}. The combination of modules through this
operation is achieved by placing a declaration of the form
\vspace{-0.15cm}
\begin{tabbing}
\qquad\=\kill
\>\verb+accumulate  M1, ..., Mk.+
\end{tabbing}
\vspace{-0.15cm}
in a new module being constructed, assuming \verb+M1+, ...,
\verb+Mk+ are module names. The lexical effect of this declaration is
to provide access to the signatures of the modules \verb+M1+, ...,
\verb+Mk+ within the new module. The dynamic semantics 
of this construct is, once again, specified by
recourse to the logical essence of modules. Let $E_1,\ldots,E_k$ be
the formulas corresponding to the modules
\verb+M1+, ..., \verb+Mk+. Then, the formula corresponding to
the new module is obtained by first conjoining $E_1,\ldots,E_k$ with
the formula obtained for the module by ignoring the accumulation and
subsequently raising the existential quantification at the head of
each of the $E$ formulas to scope over the entire conjunction; of
course, the outward movement of existential quantifiers must be done
in a logically correct way, {\it e.g.}, $(\exists x D_1(x)) \land (\exists
x D_2(x))$ should translate to $\exists y \exists z (D_1(y) \land
D_2(z))$, with $x$ in $D_1$ and $D_2$ being renamed to the distinct
and fresh variables $y$ and $z$. At a programming 
level, this logical 
interpretation is tantamount to treating accumulation as the inlining
of code in all the accumulated modules, taking care, however, to
preserve the locality of names within each.

An accumulated module could, of course, itself be accumulating other
modules. In this context, the logical interpretation of this construct
makes sense only when chains of accumulates do not close back on
themselves. The syntax of the language accordingly prohibits such
cycles.\footnote{Note that two modules that are accumulated into the
same context {\it can} utilize predicates defined in each other and
thus be mutually dependent. Prohibiting cycles in accumulation
chains is, therefore, {\it not} the same as disallowing such
dependencies.} In a separate compilation model, this is a constraint
that can only be checked at linking time.

We have at this point described the essential structure of the
modules language in its {\it entirety}: A module corresponds
syntactically to a possibly large $E$ formula obtained by combining
its clauses, its signature and the formulas corresponding to its
accumulated modules in the manner just described. The dynamic
semantics of the module is then explained {\it completely} and {\it
precisely} through this formula and the operational semantics for the
core language. The next section adds
further syntactic sugar and some compiler-oriented annotations to this
language without modifying the logical core.

%% file: modules-practical.tex
\vspace{-0.25cm}

\section{The Practical Use of Modules}\label{sec:modules-practical}

\vspace{-0.25cm}

Although the modules language described in the previous section is
simple, it is quite versatile at a programming level. We attempt to
bring this facet out in this section by considering some paradigms for
its use.  

A module in our language allows code that supports a desired
functionality to be collected into a named unit and an associated
signature provides a window into this code. Now, the
capabilities implemented by a module may be needed in the context of
another module. Module accumulation supports the realization of such
an interaction. As an example, suppose that we wish to implement a
heuristic-based graph search procedure. This procedure would
initialize a collection of states and then expand this set based on
the rules for generating new states and an underlying strategy for
selecting the next state for expansion. To realize what is required of
it, this code may need the implementation of a store. This can be
obtained by accumulating the module \verb+store+ presented earlier.
Figure~\ref{fig:graphsearch} displays part of the definition of a
graph search module to illustrate this idea. The accumulation of
\verb+store+ gives this module a type for stores that is used in the
type declarations of \verb+init_open+ and
\verb+expand_graph+. This accumulation also provides for use the
procedures \verb+init+, \verb+add+ and \verb+remove+. An aspect worthy
of note is that while the (universally quantified) variables
in the program clauses in this module can be instantiated with store
representations, these representations are still abstract: the
implicit existential quantification over local constants imposes
visibility restrictions that ensure that their inner structure can
only be accessed by recourse to operations in the module
\verb+store+. Note also that by excluding declarations
for \verb+init+, \verb+add+ and \verb+remove+ from the signature for
\verb+graph_search+, the predicate definitions in \verb+store+ can be
made entirely private to the graph search module. Thus, in contrast to
the \verb+import+ construct of \cite{Mil89jlp}, we are able to achieve
code scoping by simply structuring name spaces in a {\it completely
  statically determined} collection of code. 

\begin{figure}[t]
\begin{tabbing}
\qquad\=\qquad\qquad\=\kill
\>\verb+module graph_search.+\\
\>\verb+accumulate store.+\\
\>\verb+kind state,action type.+\\
\>\verb+type graph_search list action -> o.+\\
\>\verb+type init_open store state -> o.+\\
\>\verb+type expand_graph store state -> list state -> list action -> o.+\\
\>\verb+...+\\[3pt]
\>\verb+graph_search Soln :- init_open Open, expand_graph Open nil Soln.+\\
\>\verb+init_open Open :- start_state State, init Op, add State Op.+\\
\>\verb+expand_graph Open Closed Soln :-+\\ 
\>\>\verb+remove State Open Rest, final_state State, soln State Soln.+\\
\>\verb+expand_graph Open Closed Soln :-+\\
\>\>\verb+expand_node State NStates,+\\
\>\>\verb+add_states NStates ROp (State::Closed) NOp,+\\
\>\>\verb+expand_graph NOp (State::Closed) Soln.+\\
\>\verb+...+
\end{tabbing}
\vspace{-0.5cm}
\caption{A Module Implementing Graph Search}
\vspace{-0.5cm}
\label{fig:graphsearch}
\end{figure}

The above example shows how a {\it private} copy of code can be
acquired by a module. While this may be the desired behaviour in some
situations, the accumulated module may in many other cases represent a
common capability that is to be shared between different modules in a
large system. A specific example of this kind is provided by
``library'' predicates such as \verb+append+ on lists: given the
ubiquitous nature of lists, these predicates are likely to be needed
in many places and it is desirable to use {\it one} copy of the code
in the entire system rather than replicating the code at each place it is
needed. 

A solution to this problem is to think of modules that use such
library capabilities as being {\it parameterized} by them. Our modules 
language supports this kind of parameterization in a natural
way. For
example, suppose that we wish to think of the module \verb+graph_search+
as one that depends on an externally provided implementation of
stores rather than one that it accumulates. This dependency can be
manifest by including appropriate declarations in its signature. In
particular, this signature would identify the type constructor
\verb+store+ and the constants \verb+init+, \verb+add+ and
\verb+remove+. We would, of course, have to provide the module with
the functionality it needs eventually. This can be done by 
accumulating the module \verb+store+ at the relevant place. As a
specific illustration, suppose we wish to test our implementation of
graph search. A harness suitable for this purpose can be expressed via the
following module: 
\vspace{-0.15cm}
\begin{tabbing}
\qquad\=\kill
\>\verb+module test_graph_search.+\\
\>\verb+accumulate graph_search, store.+
\end{tabbing}
\vspace{-0.15cm}
By endowing this module with a signature that makes the needed
types, data representations and predicates from the module
\verb+graph_search+ externally visible, we can pose queries against it
that exercise the capabilities that are to be
tested.\footnote{This discussion also shows how different modules in a
composite system can share ``library'' capabilities: they can be
parameterized as described and the parameter can be
discharged by accumulating the library module into a common context.}

Predicate definitions in the logic programming context can be expanded
by adding further clauses. The definitions emanating from an
accumulated module have the potential of being extended in this way in
the accumulating module. It is often desirable 
to curtail this possibility. Referring to the module
\verb+store+, for instance, we may want to freeze
the definition of the operations \verb+init+, \verb+add+ and
\verb+remove+ that it provides. This possibility is supported by 
permitting an ``export'' annotation in
signatures. Specifically, by replacing
\vspace{-0.25cm}
\begin{tabbing}
\qquad\=\kill
\>\verb+type  add  A -> (store A) -> (store A) -> o.+
\end{tabbing}
\vspace{-0.25cm}
with the annotated declaration 
\vspace{-0.25cm}
\begin{tabbing}
\qquad\=\kill
\>\verb+exportdef  add  A -> (store A) -> (store A) -> o.+
\end{tabbing}
\vspace{-0.25cm}
in the signature of the module \verb+store+, we may signal that the
definition of \verb+add+ may not be altered by the accumulating
context. Paying attention to the model for pairing functionality that
we have just sketched, our modules language also provides a
complementary ``useonly'' annotation. Thus, by using the declaration 
\vspace{-0.5cm}
\begin{tabbing}
\qquad\=\kill
\>\verb+useonly  add  A -> (store A) -> (store A) -> o.+
\end{tabbing}
\vspace{-0.25cm}
in the module \verb+graph_search+ instead of the type declaration for
\verb+add+ we indicate that this predicate is only to be used without
alteration in this module. We note that at a logical level both the 
\verb+exportdef+ and \verb+useonly+ declarations are {\it identical
to type declarations}. They differ from type declarations only at a
pragmatic level by imposing special wellformedness restrictions---that
must be checked and can be made use of by a compiler---on 
module composition. The restrictions can, however, be quite useful in
practice: they impart a completeness property to definitions that can
help in reasoning about program properties and also in generating
better object code especially in a separate compilation setting. 

The parameterization idea that we have described requires 
signatures to be repeated in several places. To simplify program
structure and thereby to provide better documentation, we provide the
ability to accumulate signatures. To include a signature in
another signature or module, a declaration of the
form   
\vspace{-0.25cm}
\begin{tabbing}
\qquad\=\kill
\>\verb+accum_sig  foo.+
\end{tabbing}
\vspace{-0.25cm}
may be used. A wrinkle here is that the signature \verb+foo+ may
contain annotations on predicates that mark them as of an
\verb+exportdef+ kind that may actually need to be complemented in the
accumulating context. The declaration 
\vspace{-0.25cm}
\begin{tabbing}
\qquad\=\kill
\>\verb+use_sig foo.+
\end{tabbing}
\vspace{-0.25cm}
may be used instead in these cases. This declaration also causes the
mentioned signature to be included in place but only after  each 
\verb+exportdef+ declaration is converted to a \verb+useonly+ one.

While predicate definitions will usually be determined entirely by
specific modules either because of annotations of the kind described
in this section or because they pertain to local constants, they can
if needed be distributed across  
interacting modules. This feature has some uses as witnessed by
multifile declarations in Prolog. As another example, 
consider the task of implementing proof relations in different
logics. A common part to all these logics may be the treatment of
propositional rules. This treatment may be isolated in a particular
module named, say, \verb+prop_logic+. A realization of first-order
logic may then accumulate \verb+prop_logic+ and extend the predicates
defined therein. While this ability to extend definitions across
module boundaries is useful, it also raises special problems for
separate compilation as we discuss next.

%% file: sep-compilation.tex
\vspace{-0.25cm}

\section{A Separate Compilation Scheme}\label{sec:sep-compilation}

\vspace{-0.25cm}

A naive implementation of the language we have described can be
obtained by a compile-time inlining of accumulated modules. 
%This is,
%in fact, what is done by the current {\it Teyjus} compiler. 
We show in
this section that this approach can be refined into one where each
module is individually compiled and the inlining is carried out by a
later linking phase. Towards exposing the issues that have to be
addressed, we first outline the naive approach below. We then use
this context to develop the improved separate compilation scheme.

\vspace{-0.35cm}

\subsection{A Naive Implementation of Module Accumulation}

\vspace{-0.1cm}

Figure~\ref{fig:nestedaccums} presents a typical example of module
interactions that an implementation must be capable of handling. 
In this example, rather than explicitly displaying signatures, we have
marked constants as either global or local directly in the code of
modules. For simplicity, we have also elided type declarations. 

\begin{figure}[t]
\begin{center}
\begin{tabular}{lll}
\verb+module m1.+ & \verb+module m2.+ &\\
\verb+global r,w.+ & \verb+global r.+ & \\
\verb+[clauses in m1]+ & \verb+[clauses in m2]+ &\\
              &              &\\
\verb+module m3.+ & \verb+module m4.+ & \verb+module m5.+\\
\verb+accumulate m1.+ & \verb+accumulate m2.+ & \verb+accumulate m3,m4.+\\
\verb+local r.+ & \verb+local r, w.+& \verb+local q.+  \\
\verb+global w.+ & & \verb+global w.+ \\
\verb+[clauses in m3]+\hspace{50pt} &
             \verb+[clauses in m4]+\hspace{50pt} & \verb+[clauses in m5]+\\
\end{tabular}
\end{center}
\vspace{-0.35cm}
\caption{An example of nested accumulation}\label{fig:nestedaccums}
\vspace{-0.5cm}
\end{figure}

A naive implementation of our language can be obtained through a
process of inlining accumulated modules. However, this process has to
be careful about distinguishing constants that come from
different accumulated modules and must map them to
appropriately scoped ones in the larger module it constructs. A
schematic depiction of what such an inlining compiler must accomplish
appears in Figure~\ref{fig:inlined}. We have used here the
names \verb+r[1]+ and \verb+r[2]+ to distinguish the two constants
with name \verb+r+ that come from the modules \verb+m3+ and \verb+m4+
and we have similarly employed the names \verb+w[1]+ and
\verb+w[2]+ to differentiate between the global constant in module
\verb+m5+ and the local constant in module \verb+m4+ that share the
name \verb+w+.  Further, we have exploded the renaming process into a 
cascade of steps following the accumulation chain. For example, the
constant \verb+r+ appearing in the clauses of module \verb+m1+ is to
be renamed to the first local in the enclosing context (module
\verb+m3+) which is itself renamed to the second local in the
outermost context. In reality, an inlining compiler can collapse
this nesting by actually carrying out the sequence of renamings,
yielding a module with one set of global and local constants and a
collection of clauses from all the modules with the constant in them
appropriately identified. It can then proceed to compile the clauses with
complete knowledge of all the relevant predicate definitions. 

\begin{figure}[t]
\begin{center}
\begin{tabular}{l}
\verb+module m5.+\\
\verb+global w[1].+\\
\verb+local q, r[1], w[2], r[2].+  \\[3pt]
\verb+{accumulate m3 [r -> second local: r[1], w -> first global: w[1]]+\\
\verb+ global w.+\\
\verb+ local r.+\\
\verb+ [clauses from m3] with constant references suitably resolved+\\[3pt]
\verb+ {accumulate m1 [r -> first local: r, w -> first global: w]+\\ 
\verb+  [clauses from m1] with constant references suitably resolved}}+\\[5pt]
\verb+{accumulate m4 [r -> fourth local: r[2], w -> third local: w[2]]+\\
\verb+ local r, w.+\\
\verb+ [clauses from m4] with constant references suitably resolved+\\[3pt]
\verb+ {accumulate m2 [r -> first local: r]+\\
\verb+  [clauses from m2] with constant references suitably resolved}}+\\[5pt]
\verb+[clauses from m5] with constant references suitably resolved+\\
\end{tabular}
\end{center}
\vspace{-0.35cm}
\caption{The definition of module m5 after inlining
  accumulates}\label{fig:inlined}
\vspace{-0.5cm}
\end{figure}

\vspace{-0.35cm}

\subsection{A Separate Compilation Based Treatment}

\vspace{-0.1cm}

The previous model indicates the structure of the code that needs to
be produced prior to execution. We are interested, however, in
generating this code from compiled versions of each of the 
modules \verb+m5+, \verb+m4+, \verb+m3+, \verb+m2+ and \verb+m1+ that
have been generated without knowledge of where they are going to be
used and information at most of the signatures of modules that they
accumulate. In this situation
\vspace{-0.15cm}
\begin{enumerate}
\item the compiler will not have specific knowledge when compiling a
  potentially accumulated module of what the global and local constants
  are going to be mapped to in the enclosing context,

\item for predicates whose names are global and whose definitions are
  extendible, the compiler will have to produce code assuming that the 
  clauses in the module form an incomplete set and must be fitted
  into a larger context, and

\item the code that the compiler produces may have calls to predicates
  whose entry points cannot be determined at compilation time but must
  wait till the relevant assembly of modules is put together. 
\end{enumerate}
\vspace{-0.15cm}
The tasks of the inlining compiler will, under these circumstances,
have to be  divided between  a compiler that produces code for each
module separately but includes in such code suitable annotations that
allow it to be fitted into a larger context and a linker that uses the
``glue'' information with each module to build a complete bytecode
image of the system. We sketch the structure of these components below
that in combination achieve the desired result. For concreteness in
presentation, we will assume as a target low-level code that can be
run on an architecture closely related to the Warren Abstract Machine
(WAM) \cite{War83}. We assume familiarity with this machine structure
below. 

\vspace{-0.4cm}

\subsubsection{The Outcome of Compilation}

Given a module, the compiler that we envisage will produce a 
file for the linker that has the following items of information:

\vspace{-0.15cm}
\begin{enumerate}
\item A listing of the global constants that includes their names and
  other information such as their types that will be needed during
  execution. 

\item A listing of the local constants similar to that for the global
  ones but, this time, the names are not needed. 

\item A list of names for each accumulated module paired with a
  mapping from its global names (obtained from its signature) to 
  indices into either the local or global constant list for this
  module. 

\item A listing of the (indices of) externally redefinable predicates.

\item WAM-like code obtained by compiling the clauses presented in the
  module. Constant indices in this code will be indices into the lists
  in the header, to be patched up eventually by the linker. Calls to
  externally redefinable predicates also use indices into the listing
  of these predicates and will have to be filled in after the entry
  point for these has been finally determined.

\item A map from predicate names (represented by their indices) to
  their entry points in the code space. 
\end{enumerate}
\vspace{-0.15cm}
In the WAM setting, the code that is produced for individual clauses
defining a predicate is surrounded by instructions for sequencing
through choices and also indexing into them. The typical structure for
this is illustrated below:
\vspace{-0.15cm}
\begin{tabbing}
\qquad\=\verb+L11: +\=\kill
\>\>\verb+try_me_else L1+\\
\>\>\verb+switch_on_term V1,C1,Lst1,S1+\\
\>\verb+C1:+\>\verb+switch_on_constant CHT1+\\
\>\verb+S1:+\>\verb+switch_on_structure SHT1+\\
\>\verb+V1:+ \>\verb+try_me_else L12+\\
\>\> \verb+[code for one clause]+\\
\>\verb+L12:+\>\verb+retry_me_else L13+\\
\>\>\verb+...+\\
\>\verb+LLn:+\>\verb+trust_me+\\
\>\>\verb+[code for last clause]+\\
\>\verb+L1:+\>\verb+retry_me_else L2+\\
\>\>\verb+[code for another block]+\\
\>\>\verb+...+\\
\>\verb+Ln:+\>\verb+trust_me+\\
\>\>\verb+[code for last block]+
\end{tabbing}
\vspace{-0.15cm}
At the outermost level, this code captures a possible sequencing
through different chunks of clauses. Each chunk corresponds to a
subsequence over which indexing may be useful. One component of this
``indexed'' subsequence pertains to the variable case that must
support a simple sequencing through the entire collection. The other
possibility, corresponding to lists or hashing on constant name or
structure names, is that only some of the clauses in the sublist are
relevant. In this case, auxiliary sequencing code using the
instructions \verb+try+, \verb+retry+ and \verb+trust+ would be
generated. These possibilities are not specifically illustrated above
but are nonetheless relevant to the discussions that follow.

\vspace{-0.4cm}

\subsubsection{The Linking Process}

Linking begins by creating a frame for a flattened image of the
top-level module to be filled out by functions that map global
and local constants to runtime indices, recursively load the
accumulated modules and, finally, add the code to the frame. 

The mapping of constants to runtime indices follows the cascading
structure of the inlining compiler, except that this is done at
linking time. In some detail, each local constant in the chain of
accumulates translates into a unique index. The global constants of
the top-level module also are assigned unique indices. Finally, the
assignments for the local and global constants of the parent
module and the associated renaming functions determine the indices of
global constants of accumulated modules.

An issue that is important in preparing the code for addition to the
frame is that of combining predicate definitions: different modules 
may provide pieces of the definition of a predicate and these need to
be assembled together. To begin with, there must be a fixed order
governing the assembly---in $\lambda$Prolog, for instance, clauses from
the accumulated modules appear first in the order of accumulation
followed by those in the parent module---and this is adhered to by the
linker. Now, the code that is generated for a predicate in each module
has the structure of a list and a natural first step towards
integration is appending separate lists together. Ignoring for the 
moment the existence of indexing in the WAM code, it is easy to
see how this might be done. For instance, suppose that two clauses
sequences that have been compiled into the forms shown in the left and
right halves of the display below have to be combined:
\vspace{-0.15cm}
\begin{center}
\begin{tabular}{llll}
\verb+L1: +& \verb+try_me_else L2+ &\verb+L4 + &\verb+try_me_else L5+\\
& \verb+[code for a block]+\qquad\qquad & & \verb+[code for a block]+\\
\verb+L2: + &\verb+retry_me_else L3+& \verb+L5: + &\verb+retry_me_else L6+\\
& \verb+[code for a block]+\qquad\qquad & & \verb+[code for a block]+\qquad\qquad\\
\verb+L3: + &\verb+trust_me+& \verb+L6: + & \verb+trust_me+\\
& \verb+[code for a block]+\qquad\qquad & & \verb+[code for a block]+\qquad\qquad\\
\end{tabular}
\end{center}
\vspace{-0.15cm}
This combination can be realized by changing the \verb+trust_me+
  instruction that precedes the last block of the first list into a
  \verb+retry_me_else+ instruction pointing to the start of the next
definition and by changing the first instruction of that collection
  into a \verb+retry_me_else+ to yield the following:
\vspace{-0.2cm}
\begin{center}
\begin{tabular}{llll}
\verb+L1: +& \verb+try_me_else L2+ &\verb+L4 + &\verb+retry_me_else L5+\\
& \verb+[code for a block]+\qquad\qquad & & \verb+[code for a block]+\\
\verb+L2: + &\verb+retry_me_else L3+& \verb+L5: + &\verb+retry_me_else L6+\\
& \verb+[code for a block]+\qquad\qquad & & \verb+[code for a block]+\qquad\qquad\\
\verb+L3: + &\verb+retry_me_else L4+& \verb+L6: + & \verb+trust_me+\\
& \verb+[code for a block]+\qquad\qquad & & \verb+[code for a block]+\qquad\qquad\\
\end{tabular}
\end{center}
\vspace{-0.15cm}

The indexing optimization in the WAM complicates matters a little
because some elements of the top-level sequence may be indexed
blocks. If we were to simply append the top-level sequences as
suggested, the new sequence may have two adjacent blocks of this
kind. This is undesirable: it may mean, for instance, that
we end up keeping a choice point on the stack when one is not really
needed. Fortunately, it is possible to avoid this. We can 
determine if this will occur by examining the last element of the first
sequence and the first element of the second sequence. If they are both
indexed blocks, then we proceed to merge them. 

One problem to be addressed in generating a single indexed block is,
once again, that of merging two segments of code that represent
sequencing through clauses. For the ``main'' sequences
corresponding to the variable case, this can be effected as for
top-level sequences. To treat the case when these may be sequences 
realized through \verb+try+, \verb+retry+ and \verb+trust+
instructions, we augment the instruction set with two new instructions
called \verb+try_else+ and 
\verb+retry_else+. These instructions behave like \verb+try+ and
\verb+retry+ except that that they
take an additional argument that provides the address of the code to
try upon on backtracking. Suppose now that the two blocks of
sequencing code that we need to merge are the following:
\vspace{-0.4cm}
\begin{center}
\begin{tabular}{llll}
\verb+S1: + & \verb+try L1+ &\verb+S2: +& \verb+try L4+ \\
& \verb+retry L2   +\qquad\qquad & &\verb+retry L5+\\
& \verb+trust L3+& & \verb+trust L6+
\end{tabular}
\end{center}
\vspace{-0.15cm}
This merging can be realized by changing the code to the following:
\vspace{-0.15cm}
\begin{center}
\begin{tabular}{llll}
\verb+S1: + & \verb+try L1+ &\verb+S2: +& \verb+retry L4+ \\
& \verb+retry L2+ & &\verb+retry L5+\\
& \verb+retry_else L3,S2+\qquad\qquad& & \verb+trust L6+
\end{tabular}
\end{center}
\vspace{-0.15cm}
The \verb+try_else+ instruction is needed in implementing this idea in
the case where the first block corresponds to a unique clause choice. 

The other problem that needs to be dealt with in combining indexing
blocks is that of merging hash tables corresponding to constant and
structure names. This is easy to do. The compiler can actually emit
the separate tables simply as lists of pairs of constant names and
corresponding entry points to code. The linker can determine from this
which lists have to be merged and then emit the merged lists, which are 
used by the emulator to generate the hash tables.

The last aspect that the linker must resolve is the (relative) code
location for predicates that could not be finalized at compile time.
After the combining of all the clause code has been completed, a map
is available from each predicate name in our universal namespace to
the location of its definition. These addresses can now be patched in
at the appropriate places.

%% file: conc.tex
\vspace{-0.25cm}

\section{Related Work and Conclusion}\label{sec:conc}

\vspace{-0.3cm}

This paper has described a logic based interpretation of modularity in
logic programming. A natural question that arises in assessing its
contributions is the relationship of the treatment it proposes to that
in functional programming. There are obvious similarities at the
pragmatic level to the ideas of signatures and structures in Standard
ML and existential quantification in programs looks enticingly similar
to existential types that underlie hiding in functional programming
\cite{MP88toplas}. However, there is also a fundamental difference in
that logic programming is based on {\it proof search} rather than on
{\it proof normalization}. Much of the attention in this paper has
been on spelling out a coherent logical viewpoint for modularity in
logic programming and then describing how a satisfactory computational
treatment can be provided to the associated proof search.

Another relevant comparison is with work that endows
actual Prolog systems with modularity capabilities. A major concern
within such efforts (\eg, \cite{cabeza00new,sicstus,quintus}) has been
the interpretation of metalogical predicates such as {\it call} and
the treatment of declarations relating to syntax.  The focus 
on this view of modularity has significant practical relevance---{\it
e.g.}, see \cite{cabeza00new} for its importance to language
extensibility. However, this concern is orthogonal to our primary one 
here relating to name and code scoping. With regard to the handling of
names, these other approaches have been somewhat {\it ad hoc} at a
logical level, permitting the hiding of predicate names but not those
of functors and constants, a critical aspect of data abstraction. The
treatment in the {\it Mercury} language is closer to ours
pragmatically but differs in that it requires either all or none of the
constructors of a type to be hidden \cite{becket05mercury}. Moreover,
our use of existential quantifiers can lead to richer computations
that require a more sophisticated unification procedure. Unfortunately
space does not permit a fuller discussion of this issue.

We should also contrast our work with those in logic programming that
focus on a logic-oriented approach to realizing modularity. The
proposal of Sanella and Wallen \cite{SW92} that brings ideas from ML
into the logic programming setting is one example of this. The
notions of signatures and structures in this proposal once again
correspond closely to our ideas of signatures and modules. One
{\it difference} is that \cite{SW92} does not allow for a
predicate definition to be built up across different
structures/modules; such a capability has potential usefulness in a
logic programming setting as argued in
Section~\ref{sec:modules-practical}. We also note that, in our
setting, the hiding realized through signature specifications is
explained in a logic-based way.  Another proposal is that of
Miller \cite{Miller93elp} that subsumes the constructs we have used
here. Our contribution relative to this work is to demonstrate that an
entirely static subpart of it suffices to realize scoping over clause
definitions as well. Finally we mention the work of Harper and
Pfenning \cite{HP98jlc} that adapts an ML-like approach to modularity
to an LF based logic programming language but that, like
\cite{Mil89jlp}, also allows for dynamic modifications of predicate
definitions.

At an implementation level, we have had to deal with two different
issues: the treatment of scope for existentially quantified variables
and the combining of code for a given predicate that is obtained from
compiling different modules. The second issue is pertinent 
also to multifile definitions in, for instance, the SICStus
system. The solution adopted there is different at least on
the surface: compilation is done directly to core and indexing is
realized interpretively based on a data structure that is built up
incrementally as each clause is compiled \cite{Carlsson07}. It is of
interest, however, to see if aspects of that approach can be adapted
to our separate compilation setting as well.  

There are different aspects relevant to the work we have presented
here such as the logical features underlying our proposal for
modularity, the syntax chosen to support this notion, the treatment of
name scopes in compilation and computation (or, more precisely, in
unification) and the realization of separate compilation. Each of
these aspects has received consideration individually in past work,
raising the question of what precisely the contribution of this
paper is. The main novelty here, in our estimation, is in the
way we combine these different ideas to yield a logic motivated 
approach to modularity that is pragmatically useful and that has a
simple, separate-compilation based implementation.

The modules language that we have described has been implemented
within the {\it Teyjus} system. Experience relative to this system
with the approach to scoping that we have advocated has been positive:
users have adapted easily to this method from the dynamic, import
based approach that is also supported. The ideas presented in
Section~\ref{sec:sep-compilation} are part of a re-implementation of
this system whose release is imminent and that now also supports
separate compilation.

%% file: ack.tex
\vspace{-0.25cm}

\section{Acknowledgements}

\vspace{-0.25cm}

\noindent Support for this work has been provided by
the National Science Foundation under Grant No. 0429572.
Opinions, findings, and conclusions or recommendations expressed
here are those of the authors and do not necessarily
reflect the views of the National Science Foundation.

%% file: holte-nadathur.bbl
\begin{thebibliography}{10}

\bibitem{becket05mercury}
Ralph Becket.
\newblock Mercury tutorial, 2005.
\newblock Available at the URL
  \verb+http://www.cs.mu.oz.au/research/mercury/tutorial/book/book.pdf+.

\bibitem{ciao97}
F.~Bueno, D.~Cabeza, M.~Carro, M.~Hermenegildo, P.~López, and G.~Puebla.
\newblock {\em The Ciao Prolog System}, August 1997.
\newblock Reference Manual, Technical Report CLIP 3/97, School of Computer
  Science, Technical University of Madrid.

\bibitem{cabeza00new}
D.~Cabeza and M.~Hermenegildo.
\newblock A new module system for {Prolog}.
\newblock In {\em Computational Logic - CL 2000}, pages 131--148. Springer,
  2000.
\newblock LNAI Vol 1861.

\bibitem{Carlsson07}
M.~Carlsson.
\newblock Private Communication, June 2007.

\bibitem{ISO-Prolog-Modules}
International~Organization for Standardization.
\newblock Prolog. {ISO/IEC} 13211 --- {P}art 2: Modules, 2000.

\bibitem{HP98jlc}
R.~Harper and F.~Pfenning.
\newblock A module system for a programming language based on the {LF} logical
  framework.
\newblock {\em Journal of Logic and Computation}, 8(1):5--31, 1998.

\bibitem{Mil89jlp}
D.~Miller.
\newblock A logical analysis of modules in logic programming.
\newblock {\em Journal of Logic Programming}, 6:79--108, 1989.

\bibitem{Miller93elp}
D.~Miller.
\newblock A proposal for modules in $\lambda${P}rolog.
\newblock In R.~Dyckhoff, editor, {\em Proceedings of the 1993 Workshop on
  Extensions to Logic Programming}, pages 206--221. Springer-Verlag, 1994.
\newblock Volume 798 of Lecture Notes in Computer Science.

\bibitem{MP88toplas}
J.C. Mitchell and G.D. Plotkin.
\newblock Abstract types have existential type.
\newblock {\em ACM Trans. Program. Lang. Syst.}, 10(3):470--502, 1988.

\bibitem{MP89}
L.~Monteiro and A.~Porto.
\newblock Contextual logic programming.
\newblock In G.~Levi and M.~Martelli, editors, {\em Sixth International Logic
  Programming Conference}, pages 284--299. MIT Press, June 1989.

\bibitem{Nad92int}
G.~Nadathur.
\newblock A proof procedure for the logic of hereditary {H}arrop formulas.
\newblock {\em Journal of Automated Reasoning}, 11(1):115--145, August 1993.

\bibitem{NM99cade}
G.~Nadathur and D.J. Mitchell.
\newblock System description: Teyjus---a compiler and abstract machine based
  implementation of $\lambda${P}rolog.
\newblock In H.~Ganzinger, editor, {\em Automated Deduction--{CADE}-16}, number
  1632 in Lecture Notes in Artificial Intelligence, pages 287--291.
  Springer-Verlag, July 1999.

\bibitem{NT99jflp}
G.~Nadathur and G.~Tong.
\newblock Realizing modularity in $\lambda${P}rolog.
\newblock {\em Journal of Functional and Logic Programming}, 1999(9), April
  1999.

\bibitem{SW92}
D.T. Sannella and L.A. Wallen.
\newblock A calculus for the construction of modular {P}rolog programs.
\newblock {\em Journal of Logic Programming}, 12:147--178, January 1992.

\bibitem{sicstus}
{Swedish Institute of Computer Science}.
\newblock {\em SICStus {P}rolog v3 User's Manual}.
\newblock The Intelligent Systems Laboratory, PO Box 1263, S-164 28 Kista,
  Sweden, 1991--2004.

\bibitem{quintus}
{Swedish Institute of Computer Science}.
\newblock {\em Quintus {P}rolog v3 User's Manual}.
\newblock The Intelligent Systems Laboratory, PO Box 1263, S-164 28 Kista,
  Sweden, 2003.

\bibitem{War83}
D.H.D. Warren.
\newblock An abstract {P}rolog instruction set.
\newblock Technical Note 309, SRI International, October 1983.

\end{thebibliography}
